\documentclass[pre,twocolumn,amsmath,amssymb,superscriptaddress,showpacs,longbibliography,noeprint,nofootinbib]{revtex4-1}

\usepackage{color}
\usepackage{latexsym}

\usepackage{bm, mathtools}
\usepackage{graphicx}
\usepackage{graphics,epsfig,color}
\usepackage{dcolumn}
\usepackage{times}
\usepackage[english]{babel}
\usepackage[latin1]{inputenc}
\usepackage[T1]{fontenc}

\AtBeginDocument{%
    \newwrite\bibnotes
    \def\bibnotesext{Notes.bib}
    \immediate\openout\bibnotes=\jobname\bibnotesext
    \immediate\write\bibnotes{@CONTROL{REVTEX41Control}}
    \immediate\write\bibnotes{@CONTROL{%
    apsrev41Control,author="08",editor="1",pages="1",title="0",year="1"}}
     \if@filesw
     \immediate\write\@auxout{\string\citation{apsrev41Control}}%
    \fi
}%

\usepackage{cellspace}%
\usepackage{makecell}
\setcellgapes{3pt}

\def\lsim{\lesssim}
\def\gsim{\gtrsim}

\definecolor{linkcolor}{rgb}{0,0,0.6} 
\usepackage[pdftex,colorlinks=true, pdfstartview=FitV, linkcolor= linkcolor, citecolor= linkcolor, urlcolor= linkcolor, hyperindex=true,hyperfigures=true]{hyperref} 

\begin{document}

\title{On a possible nonequilibrium imprint in the cosmic background at low frequencies}

\begin{abstract}
The cosmic background radiation has been observed 
to deviate from the Planck law expected from a blackbody at $\sim$ 2.7~K at frequencies below $\sim 3$~GHz. 
We discuss the abundance of the low-energy photons from the perspective of nonequilibrium statistical mechanics. 
We propose a mechanism of stochastic frequency--diffusion, the counterpart to stochastic acceleration for charged particles in a turbulent plasma, to modify the standard Kompaneets equation.  The resulting violation of the Einstein relation allows to take advantage of low-frequency localization and finally leads to photon cooling.  The new equation predicts a frequency distribution in agreement with the absolute temperature 
measurements of the cosmic background radiation down to about 20 MHz, for which we offer here an updated compilation.
In that sense, the so called {\it space roar} we observe today is interpreted as a nonequilibrium 
echo of the early universe, and of nonequilibrium conditions in the primordial plasma more specifically.
\end{abstract}

\author{Marco Baiesi}
\affiliation{Dipartimento di Fisica e Astronomia, Universit\`a degli Studi di Padova}
\affiliation{INFN, Sezione di Padova}
\author{Carlo Burigana}
\affiliation{INAF, Istituto di Radioastronomia, Bologna}
\affiliation{Dipartimento di Fisica e Scienze della Terra, Universit\`a di Ferrara}
\affiliation{INFN, Sezione di Bologna}
\author{Livia Conti}
\email{ livia.conti@pd.infn.it }
\affiliation{INFN, Sezione di Padova}
\author{Gianmaria Falasco}
\affiliation{Complex Systems and Statistical Mechanics, Physics and Materials Science Research Unit, University of Luxembourg}
\author{Christian Maes}
\affiliation{Instituut voor Theoretische Fysica, KU Leuven}
\author{Lamberto Rondoni}
\affiliation{Dipartimento di Matematica, Politecnico di Torino}
\author{Tiziana Trombetti}
\affiliation{INAF, Istituto di Radioastronomia, Bologna}
\affiliation{CNR, Istituto di Scienze Marine, Bologna}

\maketitle

\section{Introduction}
The cosmic microwave background (CMB) is a prime witness to the physics of the early universe. According to the standard model of physical cosmology, it carries information about an epoch before neutral atoms were formed, 
dating from some $10^5$ years after the Big Bang (see \cite{1977NCimR...7..277D,sunyaevkhatri2013} for reviews). 
It is generally assumed then that matter and radiation were approximately in thermodynamic
equilibrium, owing to the high efficiency of Compton scattering, bremsstrahlung and radiative Compton
processes, with characteristic time-scales much shorter than the cosmic expansion time-scale. 
At the recombination era, when electrons and protons formed hydrogen atoms, light decoupled from matter and the 
photons started to move almost freely through the expanding universe, influenced by secondary effects only, such as the cosmological 
re-ionization associated to the formation of early stars and galaxies and the weak gravitational lensing induced by cosmic structures
(see \cite{2001PhR...349..125B,2006PhR...429....1L} for reviews and \cite{2016A&A...596A.108P,2018arXiv180706210P} for recent analyses). As a result, the
distribution function of the CMB photons at later times $t$ is supposed to follow the blackbody spectrum at an equilibrium temperature $T(t)$. 
After a delicate subtraction of the intervening astrophysical emissions, 
the cosmic background appears today very close to a blackbody radiation \cite{1990ApJ...354L..37M} at a temperature of about $2.7$~K, peaking at about 160~GHz, 
in very good agreement with the Planck spectrum from about 10~GHz up to about 600~GHz.

However, there is evidence of a systematic deviation from the Planck law of a blackbody at about 2.7~K at low frequencies, in the radio tail of the cosmic background. 
That aspect has been recently brought to attention by two independent types of observations: the CMB absolute temperature excess measured 
by ARCADE~2 \cite{Fixsen:2011} and the anomalously strong absorption of the redshifted 21~cm line from neutral hydrogen measured by EDGES 
\cite{Bowman:2018}. After consideration of possible instrumental errors and after subtracting Galactic and extragalactic sources of low-frequency radiation, a strong residual emission remains in ARCADE~2 data \cite{2011ApJ...734....6S}, 
that is much larger than predicted by 
the standard theory of CMB spectral distortions.
Consistently, the intriguing EDGES absorption profile amplitude, about 2--3 times larger than expected, could be explained by a 
much stronger background radiation with respect to standard predictions
(see \cite{Barkana:2018} for an alternative explanation assuming that the primordial 
  hydrogen gas was much colder than expected).
The scientific literature in fact abounds with experimental data from low-frequency radio surveys, some going back a long time: after subtraction for Galactic 
and extragalactic contributions, they all show an excess of soft (i.e. low-frequency) photons. A concise description of cosmic background spectrum data considered in this work is given 
in Section \ref{data}.

We repeat that the CMB spectrum theory assumes 
(near-) equilibrium
conditions, e.g.~up to the time of recombination.
The equilibrium distribution, the Planck law, is the quantum analogue of the Maxwell distribution for a classical ideal gas, and as emphasized already by the pioneers of statistical mechanics, it is the distribution to 
be typically expected as a consequence of counting with Bose statistics. For the kinetics and relaxation 
to the Planck distribution, we remind the reader in Section \ref{kom} about the Kompaneets equation \cite{Kompaneets}, 
which is used in that context.
It describes the evolution of the photon spectrum due to repeated Compton scattering off a thermal bath of non-relativistic electrons, possibly towards the equilibrium Planck distribution.   
In the Kompaneets equation, the Planck law entails zero current (in frequency space)  as the result of a 
detailed balance between diffusion and drift.  That arises from the analogue of the Einstein relation or 
the second fluctuation--dissipation relation as it is called in the (classical) Fokker-Planck equation. 
Yet, kinetically there is localization at low frequencies, as implied by the proportionality $\propto \nu^2$ of the kinetic coefficients: the number (or density) of soft photons does not change easily as the interaction with the plasma-environment is damped at low frequency.
 It is that kinetic aspect that is crucially important when (even slightly) violating the Einstein relation.
 
In this paper, we no longer assume that the universe at $t\simeq 1$ sec after the Big Bang was in thermodynamic equilibrium for the relevant degrees of freedom. Instead, in Section \ref{spec}, we propose an additional turbulent diffusion in frequency space.  Its origin is the assumed chaotic nature of the turbulent plasma, where stochastic acceleration is caused by spacetime random force fields~\cite{ulam61,sturrock66,aguer10}.  Quite independent of the detailed mechanism, the central limit theorem gives an extra contribution to the diffusivity $\propto \nu^{-1}$ (best fit, $\nu^{-1.3}$) to be added to the standard diffusive contribution $\propto \nu^2$ which has its origin in phase space calculations. For sufficiently low frequencies (actually, within the MHz--GHz range), the $\nu^{-1}$ obviously dominates and leads to photon cooling. The behavior for even much lower frequencies is unexplored today (see also Appendix \ref{app:glob_en_numb}).

Finally, in Section \ref{nonkom}, we check whether the suggested nonequilibrium change in the Kompaneets equation allows one to reproduce the main features of the observational data, in particular the observed excess in the cosmic background at low frequency.  Good agreement is remarkably easy to obtain.    The imbalance between drift and diffusion in the Kompaneets equation, resulting in what is here an effective pumping towards low frequency, 
is thus understood as a nonequilibrium feature.

While the argument is statistical, it is based on the presence of nonequilibrium dynamical activity in the primordial plasma.  The suggested mechanism is formally similar to the one for population selection in various nonequilibrium distributions, as has for example been discussed for population inversion in lasers~\cite{urna}, for kinetic proofreading in protein synthesis~\cite{hopfield} and for suprathermal kappa-distributions in space plasma~\cite{nonMax}. It may be theoretically summarized in the so called blowtorch theorem~\cite{land1975,heatbounds}: here, violating the Einstein relation and adding a low-frequency diffusion immediately leads to the abundance of soft photons.

 	Nonequilibrium effects in the early universe have not been discussed extensively so far, and details on the precise mechanism cannot be provided at this point.  So far, the origin of nonequilibrium features can only be thought to reside ultimately with gravitational degrees of freedom that have influenced the nature of light and matter as may be expected in strongly non-Newtonian regimes of gravity. Furthermore, the supposed presence of frequency-diffusion most likely leads to much slower relaxation.
        
The hypothesis that the low-frequency excess in the CMB is a nonequilibrium imprint, originating at (or ultimately before) the time of the primordial plasma, does not stand alone, though.  We mentioned already the blowtorch theorem and its relevance to population inversion. 
We are also brought to investigate such an idea by the various analogies we see with
the phenomenon of low frequency spectral power enhancement that has been observed
in a number of different nonequilibrium systems,
including disordered systems~\cite{ExPel,Cugliandolo97}, fluids~\cite{Ortiz06,CPV}, driven macromolecules \cite{pet18}, and vibrating solids \cite{ContiEtal13}. We also refer to the theoretical
models in~\cite{fal14,fal15} for a different type of population inversion.

\section{Observational framework}\label{data}

Measurements of the absolute temperature of the cosmic background are performed since the CMB discovery by~\cite{1965ApJ...142..419P} at 4.08~GHz.
In this paper, we consider the cosmic background absolute temperature data 
on the basis of the available measurements, including their quoted global errors that are related to limited experimental sensitivity, 
residual systematic effects, observed sky areas and uncertainties in foreground signal subtraction. 
We provide an updated and almost complete data compilation that
we use in the comparison with the predicted photon density. This is necessary for evaluating our theoretical model of Section \ref{nonkom}.
In order to make our analysis essentially independent of specific data selections, we typically avoid to use particular combinations of subsets of data, 
a possibility that we consider just for some comparisons (see also Appendix \ref{app:datacompilation});
an exhaustive investigation of the implications of adopting the various possible subsets of data will be performed in a further study.
In particular we exploit: 
\begin{enumerate}
	\item 
The data listed in Table~1 of the ARCADE 2 data interpretation paper~\cite{2011ApJ...734....6S}, but not the FIRAS ``condensed'' data at 250~GHz).
\item 
The data compilation from the various experiments reported in Table~1 of~\cite{2002MNRAS.336..592S} where joint constraints on early and late CMB spectral distortions were presented.
\item
The measurements by the TRIS experiment 
together with the long wavelength compilation in Table~1 reported in~\cite{2008ApJ...688...24G}.
\item
The extremely accurate measurements by FIRAS on board COBE ~\citep{1990ApJ...354L..37M,1996ApJ...473..576F}.
We take from \cite{1990ApJ...354L..37M} the measurements at the five lowest FIRAS frequencies while the results in \cite{1996ApJ...473..576F} are used above 68~GHz. 
A little rescaling is applied to the FIRAS data to account for the last absolute temperature calibration by \cite{2009ApJ...707..916F} at $T^* = 2.72548$~K.
We do not include the data by the COBRA experiment and by the analysis of molecular lines, as they fall in the same range of the much more accurate FIRAS measurements.
\item 
The recent data between 0.04~GHz and 0.08~GHz by \cite{Dowell:2018}. 
They refer to the extragalactic signal without any subtraction of the known contribution by extragalactic sources, that we perform as described below.
Note that the value adopted by the authors for the extragalactic background temperature at 408~MHz is consistent with the one in Table~1 of~\cite{2011ApJ...734....6S},
but not with the value in the subset of the older data in Table~1 of \cite{2002MNRAS.336..592S}. 
\end{enumerate}

As is well known, excluding the ARCADE~2 measurements, the averaged temperature of the data at $1$~GHz~$\lesssim \nu \lsim 30$~GHz is slightly below the FIRAS temperature determination
at $\nu \gtrsim 30$~GHz.
 On the other hand, the measurements below $\sim 1$~GHz and the excess at $\simeq 3.3$~GHz claimed by ARCADE~2 indicate a 
remarkable temperature increase in the radio tail of the background radiation. 

One should consider possible necessary corrections to the data, as other sources than CMB may have contributed.
The relevance of the accurate subtraction
clearly emerges 
in the ARCADE~2 data about the residual extragalactic emission presented in Table~1 of \cite{2011ApJ...734....6S}. 
The authors derive the extragalactic signal after the subtraction of the Galactic emission. Their residual extragalactic emission assumes the model by \cite{2008ApJ...682..223G} to describe the global contribution by unresolved extragalactic radio sources, expressed in terms of the antenna temperature 
$T_\text{ant} (\nu) = c^2/(2\nu^2k_\text{B})\, \int_{S_\text{min}}^{S_\text{max}} S \,N'(\nu,S)\,dS$.
Here $c$, $k_\text{B}$, $\nu$, $S$ and $N'(\nu,S)$ are the light speed, the Boltzmann constant, the photon frequency, the source flux density and the source differential number counts.
On the other hand, recent studies \citep{2018MNRAS.481.4548P,2018A&A...620A..74R,2014MNRAS.440.2791V,2016MNRAS.462.2934V} suggest an increase of $N'(\nu,S)$
up to a factor $\sim 3$ at $\sim 10 \mu$Jy and of a factor $\sim 1.5$ at $\sim 100 \mu$Jy with respect to the differential number counts by \cite{2008ApJ...682..223G},
likely to be ascribed to faint star forming galaxies and radio-quiet AGNs.
By simply rescaling at faint fluxes the differential number counts in \cite{2008ApJ...682..223G} by such factors, we find a larger contribution of unresolved extragalactic radio sources
(of about 30\%, when expressed in terms of antenna temperature, since only a fraction of the global contribution by unresolved extragalactic radio sources comes from sources at faint flux densities), always
to be subtracted from the signal to derive the residual extragalactic emission. 

The data we adopt in this study for the cosmic background absolute temperature are listed in Appendix~\ref{app:datacompilation} (see Table~\ref{tab:data}). 
We report the background temperatures according to quoted papers, by assuming the 
signal treatment originally performed by authors (second column). For the data where the model by \cite{2008ApJ...682..223G} 
was applied to subtract the global contribution by unresolved extragalactic radio sources, 
as for instance in \cite{2011ApJ...734....6S}, we report also the background temperature derived applying the higher subtraction described above
to account for possible higher differential number counts at faint flux densities (third column).
In the case of the data by \cite{Dowell:2018} we perform the subtraction using both the recipe by \cite{2008ApJ...682..223G} and this higher model.
See also Appendix~\ref{app:datacompilation} for further details. The data in Table~\ref{tab:data} are 
displayed in Fig.~\ref{fig:TvsFreq}; panel (a) refers to the background temperatures in the second column, panel (b) to the ones in the third column.
When compared to the quoted uncertainties, the higher subtraction, translated in equivalent thermodynamic temperature (see Table~\ref{tab:data}), 
gives appreciable changes between 0.022~GHz and 0.08~GHz, for the two TRIS measurements around 0.7~GHz, and, but only weakly, for the two ARCADE~2 measurements around 3.3~GHz.

Many explanations have been tried to account for the residual low-frequency excess, and we cannot mention all attempts.  For example, the diffuse free-free emission associated to cosmological re-ionization has been considered as one way to explain the ARCADE~2 and the radio background excess, but the signal spectral shape is steeper
than that predicted for the free-free distortion \citep{2011ApJ...734....6S}.  Furthermore, the signal amplitude is much larger than those derived for a broad set of models
(see~\cite{2014MNRAS.437.2507T,1999ApJ...527...16O}).
Efforts have also been dedicated to explain the low-frequency background signal excess and the EDGES absorption profile in terms of astrophysical emissions,
possibly in combination with particle physics phenomena
(see~\citep{PhysRevLett.107.271302,2014MNRAS.441.1147B,Barkana:2018,2018ApJ...858L..17F,2018ApJ...868...63E,2019arXiv190200511M}).

So far, there is no agreement in explaining the intriguing and still even questioned data (see~\cite{Subrahmanyan:2013,2018Natur.564E..32H,2018MNRAS.481L...6S,2019ApJ...880...26S}).
The present study is aimed at taking a very different route than previous studies, to consider the background radiation excess in the radio tail as a true cosmological signal and to explain it in terms of nonequilibrium statistical mechanics.

\section{Kompaneets equation near equilibrium}\label{kom}
The fundamental equation describing the kinetics of Compton scattering of photons and thermal electrons, 
which is relevant for the relaxation to the Planck distribution in the primordial plasma as well, 
has been introduced by Kompaneets \cite{Kompaneets}  and by Weymann \cite{weymann}.  It is assumed that the energy exchanges are non-relativistic at electron temperature $T_e$ with $k_BT_e \ll m_ec^2$ and for photon energies $h\nu \ll m_ec^2$.  Then, the dimensionless occupation number $n(t,\nu)$ at time $t$ and frequency $\nu$, obtained from the spectral energy density $E_\nu$ of the radiation via $n =c^3/(8\pi h\nu^3)\,E_\nu$, is shown to satisfy
\begin{equation}\label{kom0}
  \partial_t n= \frac{\sigma_T N_e h}{m_ec}\frac 1{\nu^2}\, \partial_\nu 
  \left\{\nu^4 \,\left[ \frac{k_\text{B} T_e}{h} \,\partial_\nu n +  (1+n)n\right]\right\} \, ,
\end{equation}
where $\sigma_T$ is the Thomson cross section and $N_e$ is the electron density. We refer to the literature~\cite{rybicki1979,katz1987} for details of the derivation of \eqref{kom0}.
The starting point is a Boltzmann equation for photons interacting with 
a plasma where the main mechanism is elastic Compton scattering between electrons and photons.  
This is thought to be the primary mechanism for the (partial) thermalization of the CMB.

We 
use a rescaled version of that Kompaneets equation,
\begin{equation}\label{kom1}
  \partial_\tau n= \frac{1}{\nu^2} \partial_\nu
  \left\{\nu^4 \,\left[ \frac{k_\text{B} T_e}{h} \,\partial_\nu n +  (1+n)n\right]\right\}
\end{equation}
for the time-evolution of the photon occupation number $n(\tau,\nu)$ with rescaled time $\tau = ht\sigma_TN_e/(m_e c)$, which is irrelevant for the stationary solution we are after.
The stationary solution of \eqref{kom1} for which the expression between square brackets vanishes, $k_\text{B} T_e \,\partial_\nu n_\text{eq} +  h\,(1+n_\text{eq})n_\text{eq}=0$, is the equilibrium Bose-Einstein distribution 
\begin{align}\label{bed}
n_\text{eq}(\nu)=\frac{1}{e^{ h \nu/(k_\mathrm{B} T_e) +C} -1}
\end{align}
which reduces to the Planck law for integration constant $C=0$ (photon chemical potential). Assuming thermal equilibrium between electrons and photons, with the usual, $\propto (1+z)$, temperature scaling for redshift $z$ due to cosmic expansion, we have $T_e (z)  = T^* (1+z)$, which is the same scaling as for the photon frequency.

In terms of the photon density (per unit frequency) defined as $\rho(\tau,\nu) := \nu^2\,n(\tau,\nu)$, with prefactor $\nu^2$ being proportional to the density of states, the equation \eqref{kom1} reads
\begin{equation}\label{kom11}
\partial_\tau \rho=  \partial_\nu
\left[ \frac{k_\text{B} T_e}{h} \,\nu^2\,\partial_\nu \rho +
\nu\left(\nu-2\frac{k_\text{B} T_e}{h}\right)\,\rho + \rho^2\right]  \, .
\end{equation}

It is the nonlinear term $\sim n^2$ in \eqref{kom1} that makes the ``low-frequency'' Rayleigh-Jeans contribution $n_\text{RJ}(\nu)\sim \nu^{-1}$ or, in \eqref{kom11},
\begin{equation}\label{rje}
\rho_\text{RJ} = \frac{k_BT_e}{h} \, \nu
\end{equation} 
for the Rayleigh-Jeans density corresponding to \eqref{kom11}.  Without that nonlinearity the stationary solution would be the Wien spectrum $n_\text{Wien}(\nu)\propto \exp [-h\nu/(k_\mathrm{B}T_e)]$, which is a good approximation for high frequencies. 
We emphasize that in all events the Planck law solves \eqref{kom1} because it balances the diffusion term (second order derivative) with the drift term (first order derivative), independent of the prefactor $\nu^4$ in front of the square bracket.  That is the usual scenario for detailed balance (or reversible) dynamics~\cite{gardiner}, for which the stationary solution shows zero current in the frequency domain.

The next important observation is the emergence of a localization effect at low frequencies, realized by the power $\nu^4$ in \eqref{kom1}.  Dynamically the escape rates away from low frequency are strongly damped, which implies for example slower relaxation for initial conditions peaking at low-frequencies.  That frequency dependence can already be read off from the Klein-Nishina cross section (for Thomson to Compton scattering).
Again, that kinetics is not visible in the equilibrium Planck distribution but it does play a role dynamically.  In fact, the low-frequency localization is a typical wave phenomena: scattering is limited at low frequencies/large wavelengths.

To be complete we note that, in the above, we considered the Kompaneets equation including only Compton scattering. The evolution equation for the photon occupation number could be described by a ``generalized'' Kompaneets equation accounting also for 
other physical processes in the plasma and coupled to an evolution equation for the electron temperature \cite{1991A&A...246...49B}. Unavoidable photon production/absorption processes operating in cosmic plasma \citep{sunyaevkhatri2013} include the double (or radiative) Compton scattering \citep{1981ApJ...244..392L,1982A&A...107...39D,1984ApJ...285..275G}, the bremsstrahlung \citep{1961ApJS....6..167K,1986rpa..book.....R} and,
in presence of primordial magnetic fields, the cyclotron process \cite{2005NewA...11....1Z}. 
In (near-)equilibrium conditions their rates are derived assuming again detailed balance and, consequently, in combination with the Compton scattering, they tend to re-establish a Planckian spectrum, as the reversible (zero current) stationary solution.  
Other photon production/absorption processes are predicted in exotic models.
Heating and cooling mechanisms not directly originating photon production/absorption can be also effectively added as source terms in the Kompaneets equation or in the evolution equation of the electron temperature,
according to a variety of almost standard or exotic processes. 
The resulting spectra mainly depend, at high redshifts, on the process epoch, the global amount of injected photon energy and number density, the overall energy exchange, 
and, at low redshifts, also on the details of the considered mechanism (see~\cite{2012MNRAS.419.1294C}).

In the following sections we neglect the effects of such additional mechanisms, focusing instead on the implication of ``violating'' the Einstein relation in the (simplest and most elementary version of the) Kompaneets equation. One must realize that the relaxation times are probably largely affected by the additional turbulent diffusion, especially in the low-frequency regime.  That can be inferred from the effective (modified) Kompaneets equation, without specifying the detailed physics.

\section{Turbulent acceleration in frequency space}\label{spec}
In what follows we consider the analogue of stochastic acceleration, but in frequency space, as a possible additional source
of momentum transfer to photons.
Stochastic acceleration is the net acceleration that particles may be expected to experience when moving under the influence of randomly spacetime-varying force fields~\cite{ulam61,sturrock66,aguer10}.   It is closely related to turbulent diffusion whereby additional diffusion in velocity space is generated due to random or chaotic accelerations.  As the net (spatially or temporally averaged) force is zero, the effect is a cumulative result on the level of most significant fluctuations, as expressed by the central limit theorem.  As an alternative picture, we can imagine particles in a nonequilibrium medium, for which the weak coupling or the Van Hove limit 
\cite{1955Phy....21..517V,1957Phy....23..441V} 
produces an additional diffusive contribution in their effective Fokker-Planck description.  In the equilibrium case, noise is accompanied with friction in an amount prescribed by the fluctuation--dissipation relation (see also next section).  In the case of stochastic acceleration, the source of noise tends to be macroscopic nonequilibrium with no or almost no compensation in terms of friction.  Without a thermal background, stochastic acceleration leads to run-away solutions as for example noted by Fermi in his explanation of the origin of cosmic (high-energy) radiation \cite{1949PhRv...75.1169F}.   Such effect, turbulent diffusion on top of thermal processes, is well-known in plasmas and has been described since the discovery of Taylor dispersion~\cite{taylor1922}. Stochastic acceleration in frequency space originates from high-frequency cancellation of energy transfer.  As an example for coupled oscillators, we refer to~\cite{eckman16,cuneo17,iubini19} where rotors with high energy tend to decouple from their neighbors due to fast oscillation of the forces. In general, the variance of the total momentum exchange scales like the inverse of the square root of the energy, as the following argument shows. 

Additionally to the thermal effects in the standard Kompaneets equation \eqref{kom1}, we consider energy transfers (over small time-interval $\sim\epsilon$) that relate momenta  $mv_i + h\nu_i/c \leftrightarrow h\nu_\epsilon/c + mv_\epsilon$ over quasi-random nonequilibrium forcing.
To be specific, from the point of view of the charged particles we suppose the presence of a random nonconservative force field $F$, e.g.~generated by electromagnetic wave turbulence. Let us suppose a classical picture where the force $F_s(x)$ changes in time $s$ at frequency $\nu_i$ while varying in space $x$ over a length $\ell$. For an electron moving at high velocity $v$ through the random medium the incurred force $G_s$ as function of time $s$ thus
decorrelates\footnote{In the sense of a persistence time for the external noise felt by the electron.} at a rate $\tau^{-1}$ which scales like $\tau^{-1}\propto \nu_i + \ell^{-1}v$.   The total momentum exchange over an arbitrarily small time $\epsilon$ is the time-integral of that force $G_s$ which means that the energy given to the photon is
\begin{align}\label{clt}
  h\nu_\epsilon - h\nu_i 
  = c\,\int_0^\epsilon ds \,G_s
  &= c\, \int_0^{\epsilon}ds\, \tilde{G}_{s/\tau}  \nonumber\\
  &=c\,\tau\, \int_0^{\epsilon/\tau}\!du\, \tilde{G}_u
\end{align}
where 
$\tilde G_{u}$ is the rescaled force field which now has persistence time of order one.
We assume that the force field is random with zero average and sufficiently ergodic.  Then, for small persistence time $\tau$, the central limit 
theorem\footnote{If $X_i$ are sufficiently independent with zero mean, $\displaystyle \lim_{n\to\infty}\ \frac 1{n}\sum_{i=1}^n X_i=0$, and finite variance, $\displaystyle \lim_{n \rightarrow \infty}\frac 1{n}\sum_{i=1}^n X_i^2=\sigma^2$,
then as $n \rightarrow \infty$, $\sum_{i=1}^n X_i = \sqrt{n}\,Z$ in distribution, where $Z$ is a Gaussian random variable with mean zero and variance $\sigma^2$.}
applies to the last integral and
\begin{equation}\label{vari}
h\nu_\epsilon - h\nu_i = c\tau\,\sqrt{\frac{\epsilon}{\tau}}\,Z = c\,\sqrt{\tau\,\epsilon}\,Z
\end{equation}
where $Z$ is a zero-mean Gaussian with finite variance containing more details about the forcing.  In particular, from \eqref{vari} and with $\tau^{-1}$ growing proportional to $\nu_i$, we expect a variance
\begin{equation}\label{bc}
\langle |h\nu_\epsilon - h\nu_i|^2\rangle \propto \frac 1{\nu_i}\, \epsilon
\end{equation}
as a function of the initial photon frequency.  In summary, by the oscillations in the electromagnetic field, time in momentum exchange gets measured in units of $1/\nu$.  As a result, small frequencies are more affected by the process. An alternative argument similar to the momentum transfer formula \eqref{clt} is obtained from inspecting time-averages of oscillating integrals.

The equation \eqref{bc} yields an additional diffusion in frequency space as the photon energy over a small time $\epsilon$ scales with $\epsilon$ and with a diffusivity which is proportional to the inverse frequency.  That brings an extra diffusion to add to the Kompaneets equation, as from the Fokker-Planck theorem
(see Section 13.3 in \cite{GrimettAndStirzaker}).
As a consequence of \eqref{bc}, we conclude by suggesting an addition to the Kompaneets equation \eqref{kom1} for $\partial_\tau n$ (see also \eqref{kom3} and the discussion at the beginning of Section \ref{nonkom}) proportional to
\begin{equation}\label{kommer}
\frac{1}{\nu^2}  \partial_\nu\left\{ \nu^2 \,\frac 1{\nu}\,\partial_\nu n\right\} \, .
\end{equation}
The $1/\nu$ is directly from \eqref{bc}, while the factors $\nu^2$ originate from the 3-dimensional Laplacian in reciprocal space with wave-vector $k$ having amplitude $|k| = \nu/c$.
We therefore write now,
\begin{align}\label{kom3}
  \partial_\tau n
  =& \frac{1}{\nu^2} \partial_\nu
  \left\{
  \nu^4 \,\left[ \frac{k_\text{B} T_e}{h} \,\partial_\nu n +  (1+n)n\right]
  \right\}\nonumber\\
  & +
  \frac{1}{\nu^2} \partial_\nu
  \left\{ \nu^2 \frac{k_\text{B} T_e}{h} \,B(\nu)\,\partial_\nu n\right\}
\end{align}
with
\begin{equation}\label{bnu1}
B(\nu) \propto \frac 1{\nu}\quad\text{  for  }\;\nu\gg \nu_1 \, .
\end{equation}
The first term on the right-hand side of \eqref{kom3} is the (original) thermal contribution \eqref{kom1} while the second part is purely diffusive and breaks the balance that before led to the Bose-Einstein equilibrium distribution \eqref{bed}: when indeed the dimensionless factor $B(\nu)/\nu^2$ depends on $\nu$, we cannot interpret \eqref{kom3} as a reversible Kompaneets equation \eqref{kom1} with a new (effectively global) temperature.

\begin{figure*}[!t] 
  \begin{tabular}{ll}
    (a) & (b) \\
    \includegraphics[width=0.48\textwidth]{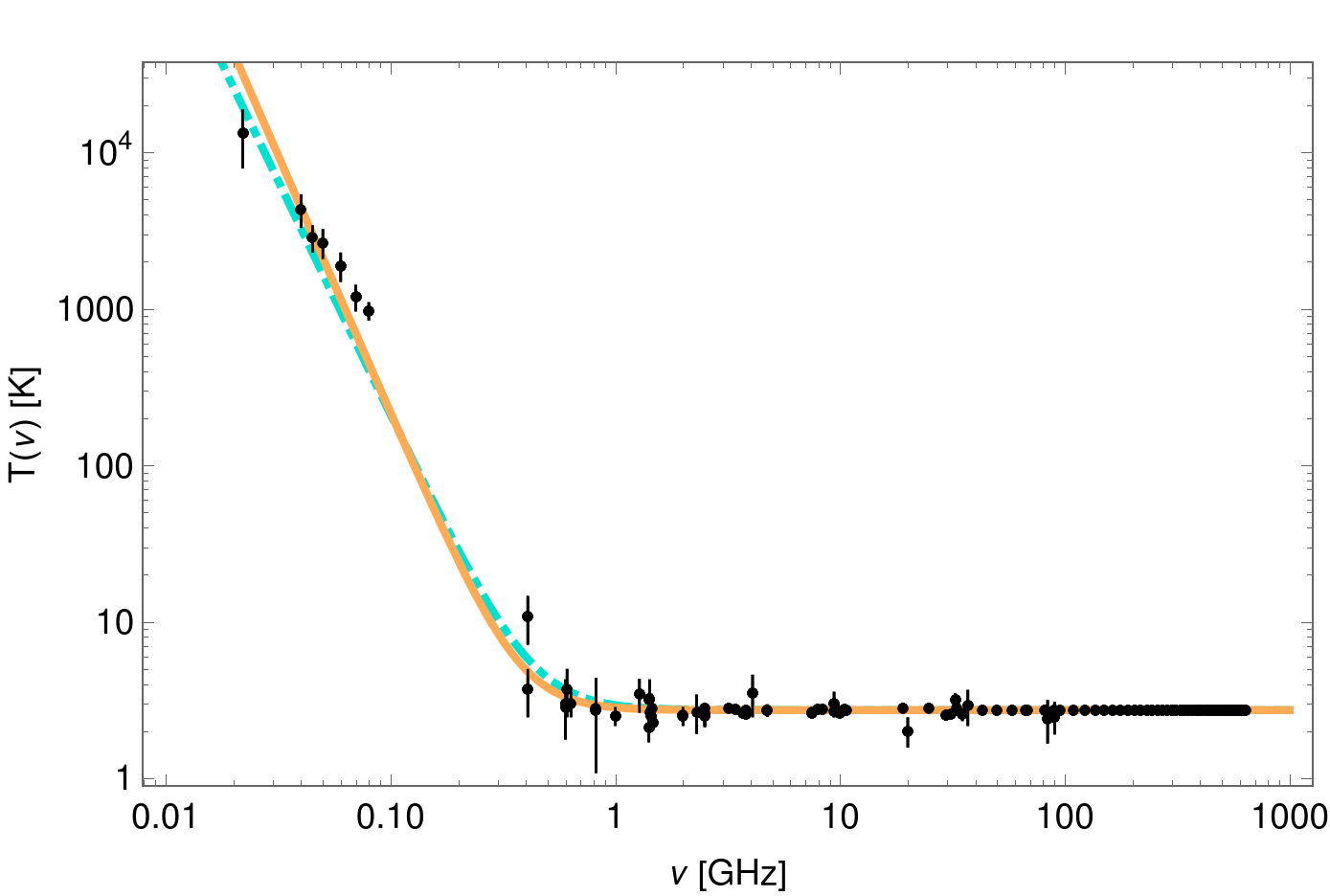} &
    \includegraphics[width=0.48\textwidth]{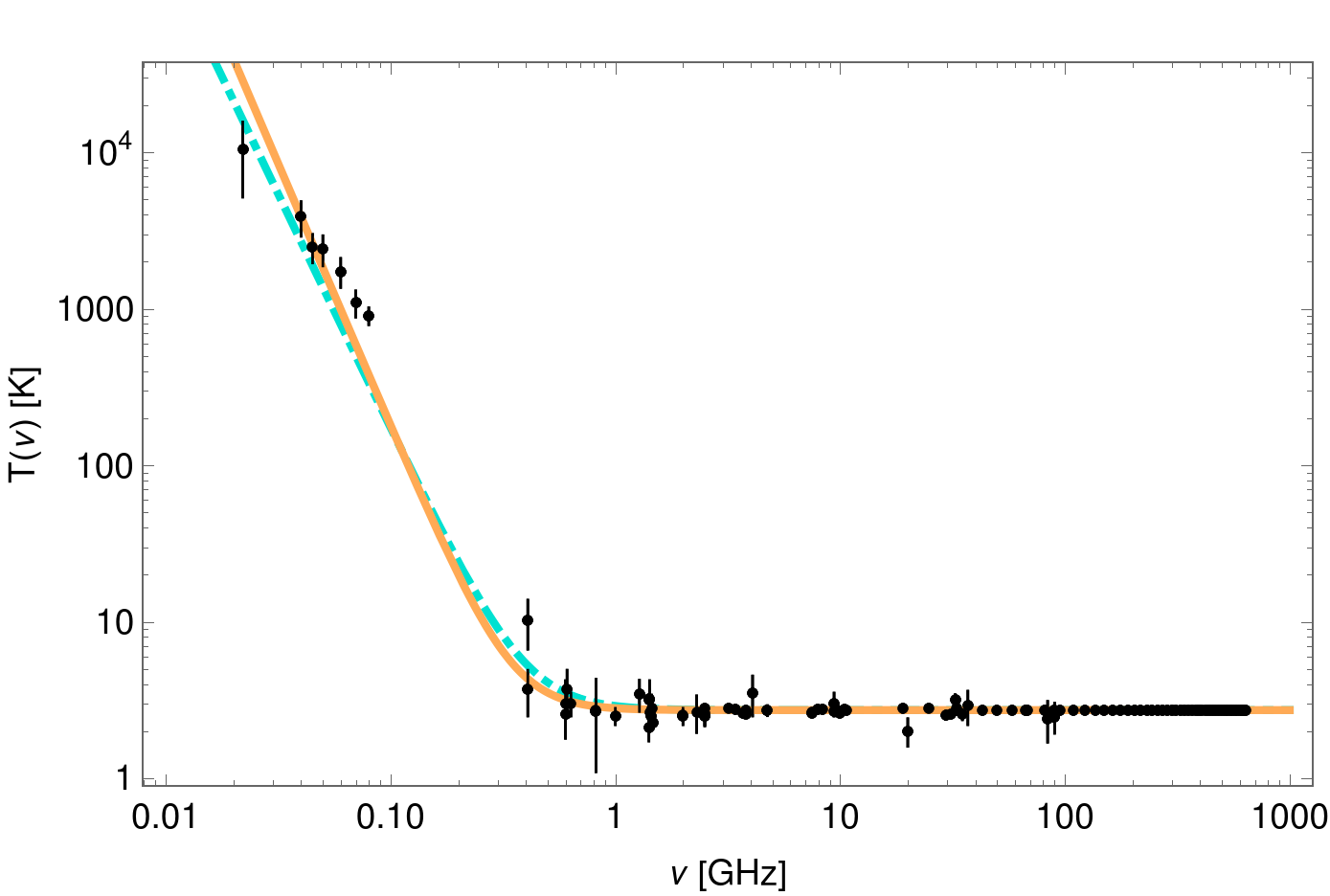}
  \end{tabular}
  \caption{Cosmic background absolute temperature as function of the frequency (in~GHz). The experimental data (black dots with $1\sigma$ error bars) refer to the measurements discussed in Section 
    \ref{data}: panel (a) refers to smaller subtraction of extragalactic signal, panel (b) refers to the higher subtraction (see Table \ref{tab:data}).
    For each panel, the turquoise dot-dashed line is the best fit of the data with $T(\nu)$ obtained from \eqref{TF}, with $\alpha=3$ and $\nu_0$ as the only free parameter: 
    for case (a) we find $\nu_0 = 0.42\pm 0.04$ GHz with a reduced $\chi^2 = 2.1$, while for case (b) we find $\nu_0 = 0.40\pm 0.05$ GHz with a reduced $\chi^2 = 2.16$ 
    (errors at 95$\%$ confidence level).
The solid orange lines are the best fits with both $\alpha$ and $\nu_0$ as free parameters: at 95$\%$ confidence level, and respectively for the cases of panel (a) and (b),
the fitting procedure results in $\nu_0 = 0.38 \pm 0.05$ GHz, $\alpha = 3.30 \pm 0.22$ with a reduced $\chi^2 = 1.91$ and 
$\nu_0 = 0.35 \pm 0.06$ GHz, $\alpha = 3.36 \pm 0.28$ with a reduced $\chi^2 = 1.93$.}
\label{fig:TvsFreq}
\end{figure*}

In \eqref{bnu1} we also indicate  a frequency-regime $\nu \gg \nu_1$ for the proposed $B(\nu) \propto 1/\nu$.  A precise quantitative value for $\nu_1$, such that the above applies for $\nu\gg \nu_1$, is difficult to  determine from the above heuristics, but the analysis of the data (in the next section, to be summarized in Figs.~\ref{fig:TvsFreq} and \ref{fig:DgBn}) suggests $\nu_1 < 10^{-2}$ GHz.

The above argument is general and powerful as it does not depend on the specific mechanisms of scattering or interaction.  It is a statistical argument based on the assumption of a sufficiently chaotic energy transfer, somewhat similar to the famous Stosszahlansatz in the kinetic theory of gases. Note, however, that a nonequilibrium input is needed: the randomness is on the level of the force $F$ in \eqref{clt} and the force is not allowed to be conservative (gradient of a potential) as a time-extensive integral (or, path-dependent work) is necessary for the application of the central limit theorem in \eqref{clt}.

The random force field $F$ appearing in the argument above is reminiscent of the intrinsic fluctuations in the CMB that were detected by COBE in the 1990s. Their origin is obviously an open problem for fundamental physics.  Yet, from a general perspective, the presence of nonequilibrium dynamical activity is perhaps not surprising at all.
Under the low-entropy assumption for the very early universe \cite{penrose-roadtoreality-2005}, 
it is not strange to believe that 
the primordial plasma was active and not starting in global thermal equilibrium. A far-from-equilibrium initial plasma would have very large relaxation times for the low frequencies.  Analogously, in space plasmas we see suprathermal tails in the electron velocity distribution, stemming from a high energy localization in the electronic degrees of freedom coupled to a turbulent electromagnetic field \cite{nonMax}.  
That may have contributed to the abundance of soft photons and together prevented thermalization before the radiation became free CMB, and a near-steady occupation (see Eq. \eqref{ns} below) was installed.
The low-frequency localization, already present in the reversible Kompaneets equation, is then the final complement to the high energy localization in the electron momenta-transfer.

Obviously, details of the mechanism will need to be added, and other scenarios may be imagined. Here we just notice that a proper quantum mechanical treatment of the interactions between photons and nonequilibrium collective plasma excitations leads in the semi-classical limit (see Eq.~(A9) in \cite{bro90}) to the dissipative Kompaneets equation \eqref{kom3}. Also, we refer to \cite{sunyaev1971} and to \cite{2015PTEP.2015g3E01T} for other examples and derivation of low-frequency distortions due to the induced Compton scattering. Another way to transfer energy from photons to electrons is to  think of pair production from high-energy photons.  Pair production in a rapidly expanding universe, such as under early inflation, will then create real long-lived high-energy electrons while depleting the high-frequency photon spectrum.

\section{Breaking the Einstein relation}\label{nonkom}

The present Section discusses the consequences of the structure \eqref{kom3}--\eqref{bnu1} that we have argued for above.  
At this moment it is instructive to consider an even larger class of Kompaneets equations. The modified Kompaneets equation \eqref{kom3} is a special case of the type
\begin{equation}\label{kom2}
  \partial_\tau n=  \frac{1}{\nu^2} \partial_\nu
  \left\{\nu^2\left[D(\nu)\partial_\nu n + \gamma(\nu) (1+n)n\right] \right\}.
\end{equation}
The notation suggests to think of $D(\nu)$ as a frequency-dependent diffusion, and of $\gamma(\nu)$ as a frequency-dependent friction. 
In \eqref{kom1} the diffusivity $D(\nu) = k_\text{B} T_e \nu^2/h $ and the friction coefficient $ \gamma(\nu)=\nu^2$ are linked by the `Einstein relation' $D(\nu)/\gamma(\nu) = k_\text{B} T_e /h$. That last property, with $D(\nu)/\gamma(\nu)$ independent of $\nu$, ensures the reversible solution \eqref{bed}.   The appearance of $\nu^2$ in the diffusivity \eqref{kom2} is an entropic effect related to the evaluation of phase space integrals in the derivation of the Kompaneets equation.  It refers to the degeneracy of the energy for a given frequency.  The real issue therefore is \eqref{bnu1} which is additive to thermal diffusion.

Comparing \eqref{kom3} with \eqref{kom1}, we retain $\gamma(\nu) = \nu^2$ for the friction, but the diffusivity changes by the addition of $B(\nu)$ for which we argued the decay \eqref{bnu1}.  For the sake of data-analysis we generalize that to the form
\begin{equation}\label{kom4}
D(\nu) =  \frac{k_\text{B} T_e}{h}\, \left[\nu^2 + \nu^{2-\alpha}\,\frac{\nu_0^{\alpha}}{\alpha +1}\, \right] 
\end{equation}
which corresponds to the dependence
\begin{equation}
\label{bnu}
B(\nu) =  \frac{\nu^2}{\alpha +1}\, \left(\frac{\nu}{\nu_0}\right)^{-\alpha}
\end{equation}
in \eqref{kom3}. The argument of the previous Section was giving $\alpha\simeq 3$, or, the diffusivity changes from a behavior $D(\nu) \propto \nu^2$ at very large frequencies to a behavior $D(\nu)\propto \nu^{2-\alpha}$ at lower frequencies, with $\alpha \simeq 3$.   The $\nu_0$ then just appears as the cross-over frequency between the behavior $1/\nu$ and $\nu^2$, as $\nu$ grows larger.

\begin{figure*}[!t]
  \begin{tabular}{ll}
    (a) & (b) \\
    \includegraphics[width=0.48\textwidth]{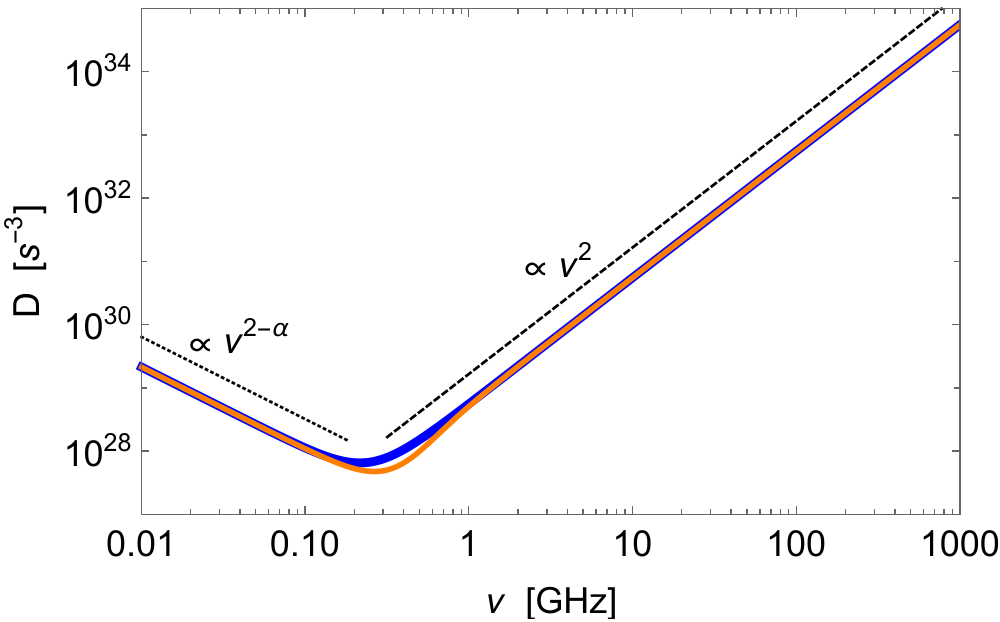} &
    \includegraphics[width=0.48\textwidth]{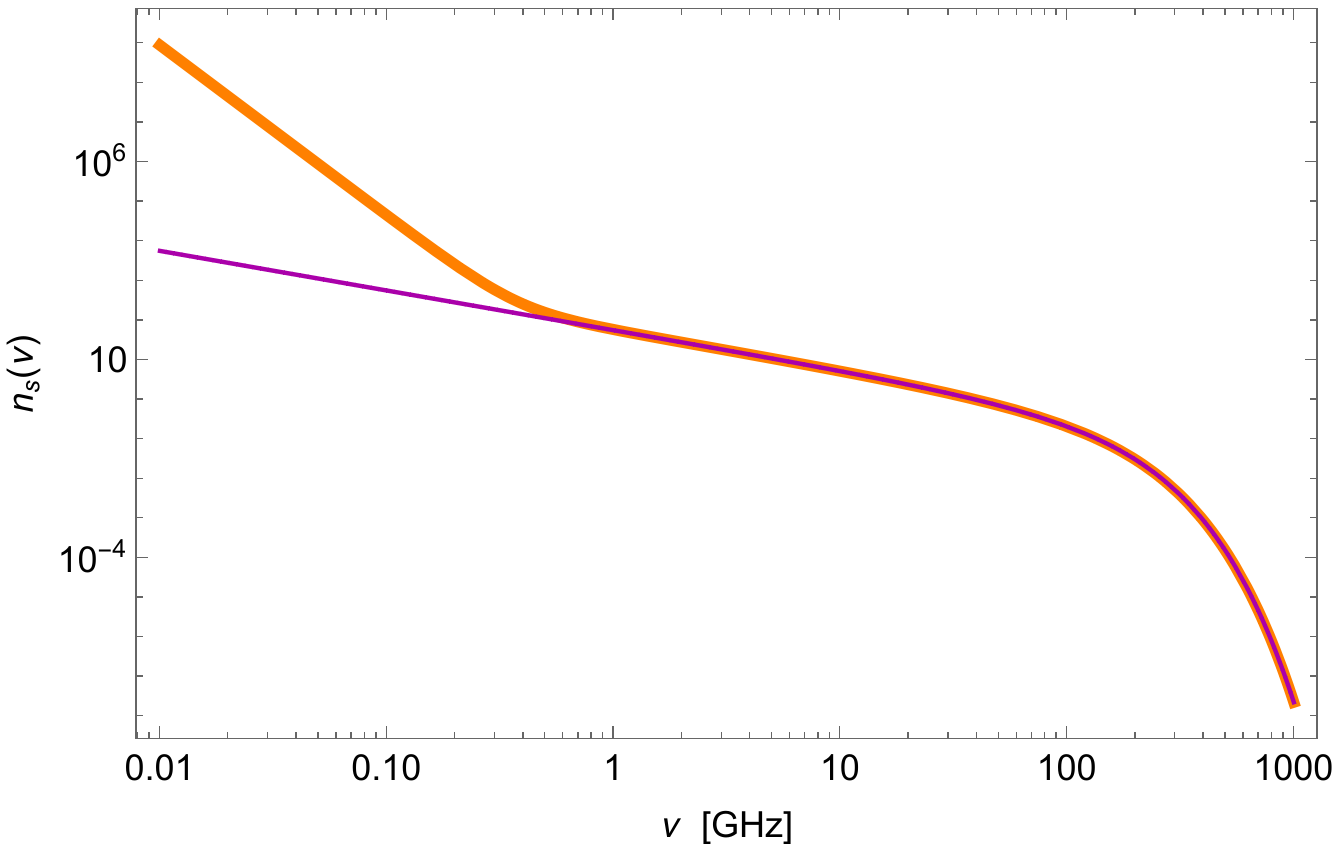}
  \end{tabular}
\caption{For the parameters resulting from the fit of the data set in column (a) of Table I with both $\alpha$ and $\nu_0$ as free
parameters. (a) The diffusivity $D(\nu)$ corresponding to the approximation in (16)
adopted in the fit (orange solid line) and, for comparison, the one
given in (13) (blue solid line); the dotted line represents a power law $\propto \nu^{2-\alpha}$ while the dashed one $\propto \nu^{2}$.
(b) Plots of the dimensionless photon occupation number $n_s(\nu)$ from and (15) and (16) (orange thick solid line) compared with the curve associated to the Planck law at $T^*$ (magenta solid line).
\label{fig:DgBn}}
\end{figure*}

Even though the Einstein relation is violated in \eqref{kom3}--\eqref{bnu}, it is quite easy to find the stationary solution of \eqref{kom3}. This is the occupation number
\begin{align}\label{ns}
  n_s(\nu)= \frac{1}{\left[\exp \int^\nu d\nu' \frac{\gamma(\nu')}{D(\nu')}\right]  -1}
  =\frac{1}{ e^{\phi(\nu)}  -1}
\end{align}
where $\phi(\nu) = \int^\nu d\nu' \gamma(\nu') / D(\nu')$ is expressed in terms of hypergeometric functions if the form (\ref{kom4}) is assumed. However, to achieve convergence when performing fits, we replace the hypergeometric function by its approximation
\begin{align}\label{F1}
  \phi(\nu) := \frac{h \nu}{k_B T_e} \;\frac{(\nu / \nu_0)^{\alpha}}{1+(\nu / \nu_0)^{\alpha}} = \frac{h \nu}{k_B T^*} \;\frac{(\nu / \nu_0)^{\alpha}}{1+(\nu / \nu_0)^{\alpha}}
\end{align}
which has the same low frequency and high frequency scalings of the exact $\phi$ in (\ref{ns}).
Since data are expressed in units of Kelvin, fits are performed with the temperature function corresponding to $\phi$,
\begin{align}\label{TF}
  T(\nu) = T_e \left[{1+\left(\frac{\nu}{\nu_0}\right)^{-\alpha}}\right]  = T^* \left[{1+\left(\frac{\nu}{\nu_0}\right)^{-\alpha}}\right] \, .
\end{align}

In \eqref{F1}, \eqref{TF} and from here onward, when referring to the comparison with observational data, we are taking $T_e$ equal to the present CMB temperature $T^*$ at high frequencies.

As our first motivation for \eqref{kom4} and making \eqref{bnu} explicit, we check whether \eqref{TF} fits the observations for $\alpha = 3$, expected on the theoretical grounds of Section \ref{spec}.  The results of the fits are shown in Fig.~\ref{fig:TvsFreq}(a) and (b) for the two data sets that we are considering, as detailed in Section~\ref{data} and listed in Table~\ref{tab:data} of Appendix \ref{app:datacompilation}. Only minimal differences in the retrieved best fit parameters are obtained for the two different radio background subtractions, without relevant changes of the whole picture. 
Overall there is a general agreement between the data and this model, with reduced $\chi^2 \simeq 2.1$ for both data sets (see also Appendix \ref{app:datacompilation}). We also perform a fit keeping both $\nu_0$ {\it and} $\alpha$ as free parameters: the results of the fits are also shown in Fig.~\ref{fig:TvsFreq}(a) and (b) for the two data sets. In this case the retrieved values are $\nu_0 \simeq 0.37$~GHz and $\alpha \simeq 3.3$, with a slightly lower
reduced $\chi^2\simeq 1.9$. This exponent is significantly larger than that found using only the data in table 1 of \cite{2011ApJ...734....6S} and is likely difficult to explain in terms of synchrotron emitters. However, this exponent is close to the value  $\alpha=3$  predicted in the previous section.
As anticipated, an informed breaking of the equilibrium assumption suffices to reproduce qualitatively the low-frequency excess observed in the data.

The diffusivity $D(\nu)$ is plotted in Fig.~\ref{fig:DgBn}(a) for $\nu_0 = 0.38 $ GHz, and $\alpha = 3.3 $.  
The effective temperature found from $k_B\,T(\nu) := h\,D(\nu)/\gamma(\nu)$ is clearly frequency-dependent,  $T(\nu)\propto \left(\nu_0/\nu\right)^{\alpha}\,T^*$ for small $\nu/\nu_0$.

The enhanced photon occupation in the low-frequency part of the spectrum is evident in Fig.~\ref{fig:DgBn}(b) where we plot the function $n_s(\nu)$, obtained by plugging in (\ref{ns}) the parameters $\alpha$ and $\nu_0$ from the fits, and the equilibrium photon occupation from the Planck spectrum. 
We should however not take \eqref{ns} as the correct behavior at ultra-low frequencies, see also the discussion in Appendix \ref{app:glob_en_numb}.

In general, in \eqref{kom2}, the term proportional to $D(\nu)\partial_\nu n$ is the transfer of (undirected) energy from the electrons (the medium) to the radiation in terms of increased intensity (number of photons). The $\gamma(\nu)$ relates to the $\nu$-dependent loss of photons. The specific breaking of the Einstein relation, the last term in \eqref{kom3}, indeed gives a noisy rate of increase of intensity with variance $B(\nu)$. The noise refers to the statistical origin of the additional diffusion, which is related to dynamical activity in the plasma. As discussed in Section \ref{spec} it is a generic effect on the level of the central limit theorem encompassing a large number of additional energy exchanges which however on average sum to zero, i.e. do not contribute to the drift.

To shed yet another light on \eqref{kom3} we write the modification of 
the Kompaneets equation \eqref{kom11} for the density $\rho$:
\begin{align}\label{kom12}
  \partial_\tau \rho=
  & \partial_\nu
  \left\{ \frac{k_\text{B} T^*}{h} \,[\nu^2 + B(\nu)]\,\partial_\nu \rho \right.\nonumber\\
  & \left.+  \left[ \nu^2-2\left(\nu+\frac{B(\nu)}{\nu}\right)\frac{k_\text{B} T^*}{h} \right]\rho + \rho^2\right\}  \, .
\end{align}

We can now make a more rigorous analogy with the Fokker-Planck equation, as we truly deal with the photon density $\rho$ 
per unit frequency: in the low-frequency approximation and by substituting \eqref{bnu}, 
we have
\begin{align}\label{kom13}
  \partial_\tau \rho =
  & \partial_\nu \left\{ \frac{k_\text{B} T^*}{h} \,B(\nu)\,\partial_\nu \rho   -2 \frac{B(\nu)}{\nu} \frac{k_\text{B} T^*}{h} \rho + \rho^2\right\}
  \\
  =
  &\partial_\nu \left\{ \frac{k_\text{B} T^*}{h}\,\frac{\nu^2}{\alpha+1}\,\left({\frac{\nu}{\nu_0}}\right)^{-\alpha}\,\partial_\nu \rho
  \right.\nonumber\\
  &\qquad \left.-2 \frac{k_\text{B} T^*}{h}\,\frac{\nu}{\alpha+1} \, \left({\frac{\nu}{\nu_0}}\right)^{-\alpha} \, \rho + \rho^2\right\}  \, . \nonumber
\end{align}
The (nonequilibrium) insertion of 
$B(\nu) \propto \nu^{2-\alpha}$ 
increases the diffusion constant for small frequencies, but there is also negative friction for small frequencies via the term 
$B(\nu)/\nu \sim \nu^{1-\alpha}$.  
The amplitude of the nonlinear term $\propto \rho^2$ is unchanged of order one, which reflects the essential localization as it derives from $\gamma(\nu)=\nu^2$.  The stationary solution of \eqref{kom13} is the modified Rayleigh-Jeans law (low-frequency regime in nonequilibrium),
\begin{align}\label{noneqRJ}
\rho_\text{mRJ}(\nu) = \frac{k_BT^*}{h} \, \nu \, \left( { \frac{\nu}{\nu_0} } \right)^{-\alpha} \, .
\end{align}
As we remember that $\alpha\simeq 3$, that shows of course a drastic increase of the density with respect 
 to the usual (low-frequency regime in equilibrium) Rayleigh-Jeans case \eqref{rje} where $\rho_\text{RJ}(\nu) = k_BT^*\,\nu/h$.

\section{Conclusions and outlook}
While the ultimate trigger of the additional photon intensity at low frequencies is arguably to be found in the original plasma, in the epoch from the quark to the hadron age of the universe, we have not considered here essential modifications to the usual Compton scattering theory between photons and electrons or to photon interactions. Instead of searching for more subtle aspects of QED as candidates to clarify the puzzling appearance of the space roar, our arguments have been statistical and kinetic.

We have modified the Kompaneets equation within an effective nonequilibrium scenario by introducing a frequency-dependent
diffusion. That leads to a violation of the Einstein relation and of the balance between diffusion and friction.  The result is a clear enhancement of lower photon frequencies compatible with the best data available for the cosmic background radiation. One crucial ingredient is already present in the (reversible) Kompaneets equation: the low-frequency localization.  
The other ingredient is stochastic acceleration in frequency-space as the result of nonequilibrium dynamical activity. Here a statistical argument (central limit theorem) applies up and above all details on fundamental interactions. It implies an extra source of diffusion where the diffusivity is inversely proportional to the photon frequency.  It is the combination of the low-frequency localization and that turbulent diffusion that creates a (new) stationary frequency distribution for the (nonequilibrium) Kompaneets equation. 
As a summary, the argument of stochastic frequency-acceleration has yielded the modification of the Kompaneets equation,
\begin{align}\label{komfin}
\partial_\tau n= &\frac{1}{\nu^2} \partial_\nu
\left\{ \nu^4 \,\left[ \frac{k_\text{B} T^*}{h} \,\partial_\nu n +  \,(1+n)n  \right] \right\}
\nonumber\\
&+
\frac 1{\nu^2} \, \partial_\nu\left\{ \nu^2 \frac{k_\text{B} T^*}{h} \frac{\nu_0^\alpha}{\left( \alpha+1 \right) \, \nu^{\alpha-2}}\,\,\partial_\nu n\right\}
\end{align}
to be applied in the whole frequency range $\nu > 20$ MHz, and where $\nu_0$ and $\alpha \simeq 3$ are the only fitting parameters. We have tested our theory by calculating the resulting frequency-dependent (effective) temperature of the cosmic background in a very wide range, including frequencies where excess is observed, achieving a reasonable agreement with the whole dataset for $\nu_0 \simeq 0.3-0.4$ GHz, and remarkably even when fixing $\alpha = 3$.

What seems mandatory for future explorations is an experimental effort devoted at more precise estimates of the cosmic background in the low-frequency tail,
from about $(10-20)$~GHz downward. The frequency region between $0.1$~GHz and $0.4$~GHz is of particular relevance because no experimental data are available. 
Observations at frequencies even lower than so far performed seem to be 
important to test or to complement our picture, since they could reveal 
larger deviations from the blackbody radiation or the transition to 
regimes (expected towards zero frequency) different to the one explored in this work.
On the other hand, the background temperature increase predicted by our model is already significant,  
having an amplitude comparable or larger than those produced by unavoidable mechanisms with typical parameters, 
at frequencies between a few GHz and $(10-20)$~GHz, a region where foreground mitigation is likely less critical and extremely accurate observations with space missions are in principle feasible.
Thus, verifications of our model could take advantage from the next generation of both radio facilities and CMB dedicated projects. It would also be interesting to study the isotropy of the low frequency excess, since the present approach neglects this issue.  
Finally, a more accurate comprehension of Galactic and extragalactic intervening astrophysical emissions is necessary.

 From the present analysis we  conclude that low-frequency data may be evidence for important nonequilibrium features in the early universe, when quantum and gravitational effects were strongly influenced by special 
 (e.g.~low entropy) conditions  at the time of the Big Bang.

\acknowledgments

CB and TT acknowledge partial support from the INAF PRIN SKA/CTA project 
FORECaST and the ASI/Physics Department of the University of Roma-Tor 
Vergata agreement n. 2016-24-H.0 for study activities of the Italian 
cosmology community. TT acknowledges partial support from the research 
program RITMARE SP3 WP3 AZ3 U02 and the research contract SMO at 
CNR/ISMAR.
MB acknowledges support from Progetto di Ricerca Dipartimentale BIRD173122/17.
LR acknowledges partial support from MIUR grant Dipartimenti di Eccellenza 2018-2022.
LR is grateful to Xiamen University for unique hospitality, 
when part of this paper was written.
The authors of this publication have chosen to appear in alphabetic order.

\bibliography{references3}

\appendix

\section{Data compilation and fit procedure}
\label{app:datacompilation}

In Table \ref{tab:data} we report the data compilation described in Section~\ref{data}. As discussed there, for the data where the model by \cite{2008ApJ...682..223G} 
was applied to subtract the global contribution by unresolved extragalactic radio sources, we considered also a higher subtraction 
to account for possible higher differential number counts at faint flux densities. These two somewhat different subtractions are also applied to the radio background data by \cite{Dowell:2018}.
For the other data sets we keep the original foreground treatments performed by the authors, the differences between the two above 
extragalactic foreground subtraction models being in any case much smaller than the quoted uncertainties.

We perform our fit first with a 2-dimensional grid in $\nu_0$ and $\alpha$, to explore the dependence of the $\chi^2$ on parameters and to avoid a possible wrong convergence; to overcome the finite sampling of the grid method, we then use a nonlinear minimization tool, weighting data with their inverse squared error. The fit is achieved with a Levenberg-Marquardt method from initial values $\nu_0^{\textrm{ini}} = 0.5$~GHz, $\alpha^{\textrm{ini}} = 4$. With significantly different initial values, the convergence of the algorithm is compromised and the final result may be easily discarded basing both on visual inspection and on the results of the grid method. Fit errors are extracted from the parameters confidence interval, with a default $95\%$ confidence level.
The reduced $\chi^2 \simeq 1.9$ we found for our model reflects the use of the (almost) complete available sets of data. Indeed such a (relatively high) value is not surprising, since
different data sets are affected by different systematic effects and derived with different foreground treatments. Moreover we are assuming just a simple model as in \eqref{TF}.

\begin{figure}[!t] 
  \includegraphics[width=0.48\textwidth]{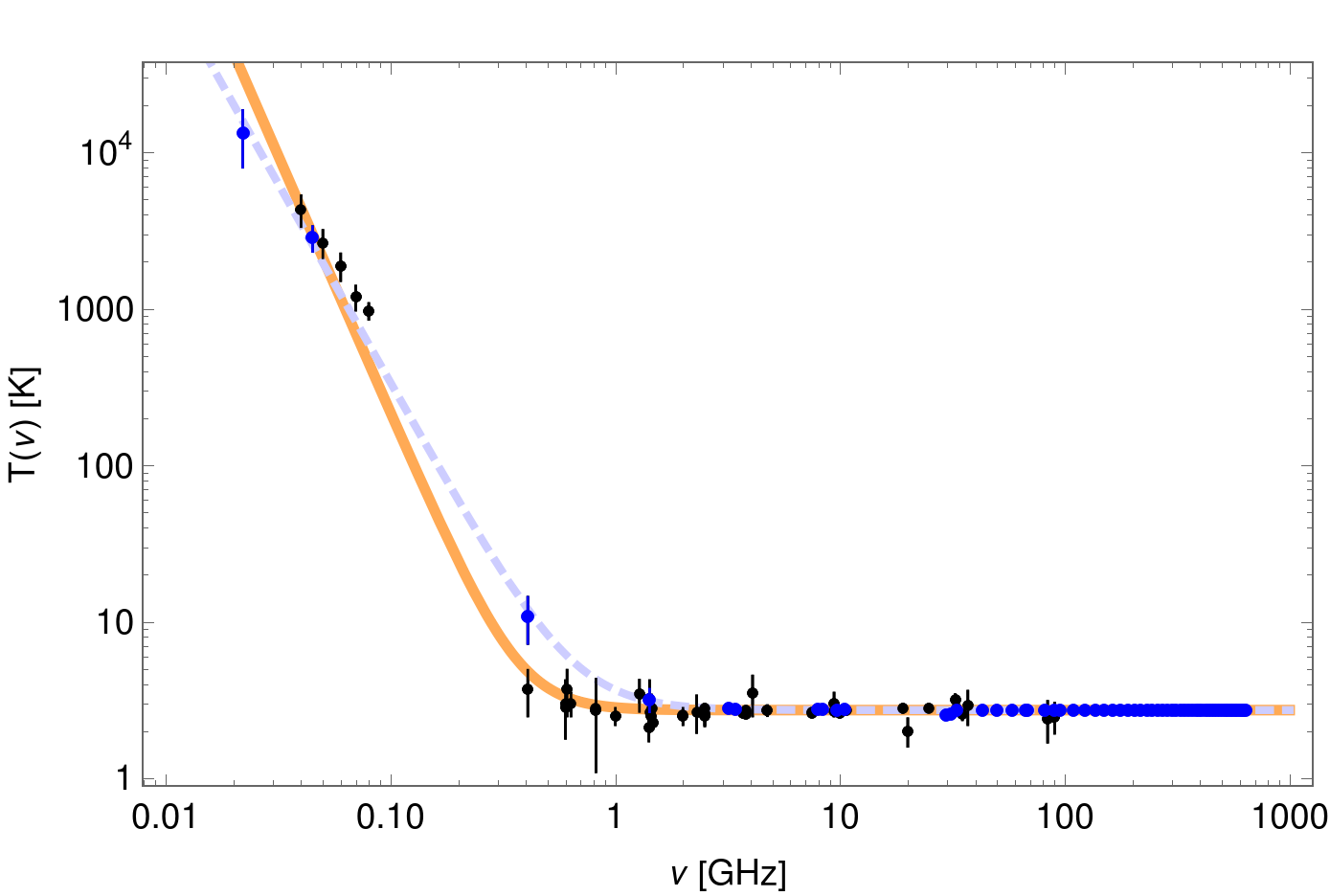}
  \caption{Cosmic background absolute temperature as function of the frequency (in~GHz).
    The experimental data discussed in the main text (with smaller subtraction of extragalactic signal) are shown as blue and black dots with $1\sigma$ error bars. The data subset from~\cite{2011ApJ...734....6S} with full set of FIRAS data is shown as blue dots.
    The light blue dashed line is the best fit to the latter (see text for the inferred parameters). For comparison we plot here also the best fit of the full dataset (orange solid line) which is discussed in the main text.
\label{fig:compar}}
\end{figure}

For an immediate comparison (and cross-check), we shall now consider a smaller data set, i.e. only the data in 
Table 1 of \cite{2011ApJ...734....6S}, but using the full set of FIRAS data and not replacing it with the ``condensed'' FIRAS value at 250~GHz. In this case we obtain a reduced $\chi^2$ of $\simeq 1.08$ with best fit parameters
(and again errors at 95\% confidence level)
$\nu_0 = 0.66\pm 0.06$~GHz and $\alpha= 2.55\pm 0.10$ 
(implying a power-law amplitude of 18.72~K at 0.31~GHz, formally in terms of equivalent thermodynamic temperature, see \eqref{TF}),
fully consistent within errors with those found in Table~2 of \cite{2011ApJ...734....6S} for the power-law fit model, as expected. In Fig.~\ref{fig:compar} we compare our best fit of the smaller data set with the best fit of the full data set which is discussed in the main text.

Applying the higher extragalactic subtraction, we find
a similar reduced $\chi^2$ ($\simeq 1.07$), $\nu_0 \simeq 0.64\pm0.06$~GHz and, as expected, a slightly smaller value of $\alpha$ ($\simeq 2.52 \pm 0.11$). Finally, for both the two extragalactic subtraction models, replacing the full set of FIRAS data with the ``condensed'' FIRAS value at 250~GHz we find similar best fit values, but with a reduced $\chi^2 \simeq 1.6$ 
in agreement with the one found in \cite{2011ApJ...734....6S}.

This simple comparison between the results found using two different data set compilations underlines the relevance of a significant improvement of both background observations and foreground modeling.

\section{Global photon energy and number density}
\label{app:glob_en_numb}

In this Appendix, we discuss the frequency range validity for the assumed $D(\nu)$ or $\phi(\nu)$, from \eqref{bnu1}, in the modified Kompaneets equation \eqref{kom3}; 
see \eqref{kom4}--\eqref{F1} (and also \eqref{bed}).

We rewrite the (nonequilibrium stationary) photon occupation number $n(\nu)$ as
\begin{align}
  n(\nu) &=  \frac{1}{e^{\phi(\nu)}-1} \nonumber\\
  & = n_P(\nu) + [n(\nu) - n_P(\nu)] \nonumber\\
  & = n_P(\nu) + \delta n(\nu) \, 
\end{align}
\noindent
where $n_P(\nu) = 1 / (e^{x_e} -1)$ is the Planckian distribution, $x_e = h \nu/(k_\mathrm{B} T_e)$  and $\delta n(\nu)$ defines the departure of $n(\nu)$ from it. 
To calculate $\delta n(\nu)$ at low frequencies, where the excess is more relevant, 
we can rely on the Rayleigh-Jeans approximation
\begin{align}
  \delta n(\nu)
  & \simeq  n(\nu)^{RJ} - n_P(\nu)^{RJ} \nonumber\\
  & \simeq \frac{1}{x_e+C(x_e)} - \frac{1}{x_e} \nonumber\\
  & = x_{e,0}^\alpha x_e^{-(\alpha+1)} \, ,
\end{align}
simplifying the computation of the global photon energy and number density:
\begin{align}
  E_r & = 8\pi \frac{(k_\mathrm{B} T_e)^4}{(h c)^3}  \int_0^\infty n(x_e) x_e^3 dx_e \nonumber\\
  & \simeq E_P + 8\pi \frac{(k_\mathrm{B} T_e)^4}{(h c)^3}  \int_{x_a}^{x_b} \, x_{e,0}^\alpha \, x_e^{-(\alpha+1)} \, x_e^3 \, dx_e \, ,
\end{align}
\begin{align}
  N_r
  & = 8\pi \frac{(k_\mathrm{B} T_e)^3}{(h c)^3}  \int_0^\infty n(x_e) x_e^2 dx_e  \nonumber\\
  & \simeq N_P + 8\pi \frac{(k_\mathrm{B} T_e)^3}{(h c)^3}  \int_{x_a}^{x_b} \, x_{e,0}^\alpha \, x_e^{-(\alpha+1)} \, x_e^2 \, dx_e \, .
\end{align}
\noindent 
Here $E_P = aT_e^4$ and $N_P = (I_2/I_3)(aT_e^3/k_\mathrm{B})$ are the global photon energy and number density for the Planckian distribution, $a = 8\pi I_3 k_\mathrm{B}^4 / (hc)^3$,
$I_m = \int_0^\infty x^m [e^x-1]^{-1} dx = m! \zeta(m+1)$ ($I_2 \simeq 2.404 $, $I_3 = \pi^4/15$),
$x_{e,0} = h \nu_0/(k_\mathrm{B} T_e)$, and $x_a$, $x_b$ (with $x_a \ll x_{e,0} \ll 1 \lsim x_b$) define the integration interval in $x_e$.
Let us write
\begin{equation}\label{E1}
E_r \simeq E_P \cdot \tilde{f} (x_{e,0},x_a,x_b,\alpha) 
\end{equation}
\noindent
and
\begin{equation}\label{N1}
N_r \simeq N_P \cdot \tilde{\varphi} (x_{e,0},x_a,x_b,\alpha) \, .
\end{equation}

\noindent
For $\alpha = 3$, we get 
\begin{equation}\label{Eeq3}
\tilde{f} (x_{e,0},x_a,x_b,\alpha) = 1 + (15/\pi^4) x_{e,0}^\alpha \log(x_b/x_a) \, ,
\end{equation}
\noindent
while for $\alpha \ne 3$, we have 
\begin{equation}\label{Enon3}
\tilde{f} (x_{e,0},x_a,x_b,\alpha) = 1 + \frac{15}{\pi^4} \frac{x_{e,0}^\alpha}{\alpha-3} (x_a^{-(\alpha - 3)} - x_b^{-(\alpha - 3)})\, .
\end{equation}
\noindent
For $\alpha = 2$, we get  
\begin{equation}\label{Neq2}
\tilde{\varphi} (x_{e,0},x_a,x_b,\alpha) = 1 + (1/I_2) x_{e,0}^\alpha \log(x_b/x_a) \, ,
\end{equation}
\noindent
while for $\alpha \ne 2$, we have  
\begin{equation}\label{Nnon2}
\tilde{\varphi} (x_{e,0},x_a,x_b,\alpha) = 1 + \frac{1}{I_2} \frac{x_{e,0}^\alpha}{\alpha-2} (x_a^{-(\alpha - 2)} -  x_b^{-(\alpha - 2)})\, .
\end{equation}

For $\alpha < 3$ (or $\alpha < 2$), we could in principle set $x_a \rightarrow 0$ in the calculation of $E_r$ (or of $N_r$), but the result depends also on $x_b$, and, obviously, $\delta n(\nu)$ could become appreciable 
also at relatively larger $x_e$ for decreasing $\alpha$, possibly requiring to go beyond the Rayleigh-Jeans limit for a precise calculation. 

Conversely, for $\alpha = 3$ 
\eqref{Eeq3} gives $(\tilde{f} - 1) \simeq - (15/\pi^4) x_{e,0}^\alpha \log(x_a)$, while for $\alpha = 2$ \eqref{Nnon2} gives
$(\tilde{\varphi} - 1) \simeq - (1/I_2) x_{e,0}^\alpha \log(x_a)$, implying a formal divergence for $x_a \rightarrow 0$.
Analogously, considering that $x_a \ll x_b$, 
\eqref{Enon3} gives 
$(\tilde{f} - 1) \simeq (15/\pi^4) [x_{e,0}^\alpha/(\alpha-3)] x_a^{-(\alpha - 3)}$ 
for $\alpha$ sufficiently larger than 3, while 
\eqref{Nnon2} gives 
$(\tilde{\varphi} - 1) \simeq (1/I_2) [x_{e,0}^\alpha/(\alpha-2)] x_a^{-(\alpha - 2)}$
for $\alpha$ sufficiently larger than 2;
in general, $\alpha > 3$ (or $\alpha > 2$) implies again a divergence of $E_r$ (or of $N_r$) for $x_a \rightarrow 0$.
More physically, $n(x_e$) should have a substantial flattening at $x_e$ below a certain dimensionless frequency $x_a$ (or at $\nu$ below a present time frequency $\nu_a$).

The relative difference of the global photon energy density with respect to the Planckian case, $\delta E_r / E_P = (E_r - E_P)/E_P \simeq \tilde{f} - 1$, is less than a certain value $\epsilon$ ($\ll 1$) for 
$x_a \gsim \exp [(15/\pi^4) x_{e,0}^{-3}\epsilon]$ if $\alpha = 3$ or for 
$x_a \gsim [(15/\pi) (x_{e,0}^\alpha / \epsilon) / (\alpha - 3)]^{1/(\alpha - 3)}$ if $\alpha > 3$.
Analogously, for $\alpha > 2$, the relative difference of the global photon number density with respect to the Planckian case,
$\delta N_r / N_P = (N_r - N_P)/N_P \simeq \tilde{\varphi} - 1$, is less than $\epsilon$ for $x_a \gsim [(1/I_2) (x_{e,0}^\alpha / \epsilon) / (\alpha - 2)]^{1/(\alpha - 2)}$.

The requirement of a change in the redshift of matter-radiation equivalence less than $\sim 1 $\%, comparable to the accuracy set by
{\it Planck} \citep{2018arXiv180706209P}, {\it i.e.} $\epsilon \sim 10^{-2}$ (a condition stronger than that set by standard cosmological nucleosynthesis), in the case of the best fit values of $\nu_0$ and $\alpha$ found in Section \ref{nonkom}
implies $x_a \gsim 2.3 \cdot 10^{-14}$, corresponding to $\nu_a \gsim 1.3 \cdot 10^{-3}$ Hz, is certainly not stringent. For comparison,
a much stronger condition $\delta E_r / E_P \lsim 10^{-5}$ (not to be confused with the potential limits on spectral distortion parameters from analyses in the near-equilibrium approach
usually performed at higher frequencies) requires $x_a \gsim 2.3 \cdot 10^{-4}$, corresponding to $\nu_a \gsim 0.013$ GHz, a value approaching the minimum frequency of current cosmic background observations.
In the case $\alpha =3$, for any significant value of $\epsilon$, we find instead $x_a$ larger than a value always negligible in practice, as expected from continuity  with the case $\alpha  < 3$.

\begin{table*}
  \scalebox{0.8}{
\begin{tabular}{@{}|c|c|c|c|c|@{}}
\toprule
$\nu$ (GHz) & $T$ (K) & $T$ (K) &  $1 \sigma$ error (K) & References \\
\hline
& (a) & (b) & (c) & From Table 1 in \cite{2011ApJ...734....6S}\\
&&&& (without``condensed'' FIRAS at 250~GHz) \\
\hline
     0.022  & 13268 & 10411 & 5229 &  From \cite{1999AAS..137....7R} 
     \\
     0.045   &  2843  & 2477  &      512               &  From \cite{1999AAS..140..145M} 
     \\
      0.408    &   10.80   & 10.21 &     3.53            &  From \cite{1981AA...100..209H} 
      \\
       1.42     &  3.181  & 3.167 &      0.526            &  From \cite{1986AAS...63..205R} 
       \\
       3.2  (d)    & 2.7770    & 2.7759   &    0.010            & From ARCADE 2 \cite{Fixsen:2011,2011ApJ...734....6S} \\
       3.41 (d)     &  2.7610 & 2.7607 &       0.008            &  "  \\
       7.98    &  2.761  & 2.760 &     0.013            &  "  \\
       8.33     &  2.742  & 2.742 &      0.015            &  "  \\
       9.72     &  2.73  & 2.73 &      0.005            &  "  \\
       10.49    &   2.738  & 2.738 &      0.006            &  "  \\
       29.5     &  2.529  & 2.529 &      0.155            &  "  \\
       31      &  2.573  & 2.573 &      0.076            &  "  \\
       90      & 2.706  & 2.706 &      0.019            &   "  \\
\hline       
& & & & From compilation in Table 1 of \cite{2002MNRAS.336..592S}\\
&&&&(years 1965--1975) \\
\hline
        0.408    &       3.7    & &      1.2  & From \cite{1967Natur.216..753H} \\
        0.610    &       3.7    & &      1.2  & From \cite{1967Natur.216..753H} \\
        0.635    &       3.0   & &       0.5  & From \cite{1970AuJPh..23..529S} \\
        1      &        2.5   & &       0.3  & From \cite{1969SvA....13..223P} \\
        1.42     &       3.2   & &       1.0  & From \cite{1967AJ.....72S.315P} \\
        1.44    &        2.5   & &       0.3  & From \cite{1969SvA....13..223P} \\
        1.45     &       2.8   & &       0.6  & From \cite{1966Natur.210.1318H} \\
        2     &         2.5   & &       0.3  & From \cite{1969SvA....13..223P} \\
        2.3     &        2.66  & &       0.7  & From \cite{OtoshiStelzreid1975} \\
        4.08     &       3.5   & &       1.0  & From \cite{1965ApJ...142..419P} \\
        9.4    &         3.0   & &       0.5  & From \cite{1966PhRvL..16..405R} \\
        9.4      &       2.69   & &      0.185 & From \cite{1967PhRvL..19.1199S} \\
        19     &        2.78    & &     0.145 & " \\
        20      &       2.0   & &       0.4  & From \cite{1967PhRvL..18.1068W}\\
        32.5     &       3.16   & &      0.26  & From \cite{1967PhRvL..19.1251E} \\
        35    &        2.56    & &       0.195 & From \cite{1967PhRvL..19.1195W} \\
        37     &        2.9    & &       0.7  & From \cite{1968SvA....11..905P} \\
        83.8     &       2.4    & &      0.7  & From \cite{1971SvA....15...29K}\\
        90      &       2.46   & &      0.42 & From \cite{1968PhRvL..21..462B} \\
        90     &        2.61   & &       0.25  & From \cite{1971PhRvL..26..919M} \\
        90      &       2.48   & &       0.54  & From \cite{1974Natur.247..528B} \\
        \hline
        & & & & From compilation in Table 1 of \cite{2002MNRAS.336..592S}\\
        &&&&(years 1985--2000) \\
\hline
 0.6      &       3.0   & &     1.2   & From \cite{1990ApJ...357..301S} \\
 0.82      &      2.7    & &    1.6   & From \cite{1991ApJ...378..550S} \\ 
 1.28      &      3.45   & &    0.78    & From \cite{2000JApA...21....1R} \\
 1.41      &      2.11   & &    0.38   & From \cite{1988ApJ...334...14L} \\
 1.425      &     2.65   & &    0.315    & From \cite{1996ApJ...458..407S} \\
 1.47      &      2.26   & &    0.19   & From \cite{1993ApJ...409....1B} \\
 2       &       2.55   & &     0.14   & From \cite{1994ApJ...424..517B}\\
 3.8      &       2.64   & &    0.07   & From \cite{1991ApJ...381..341D}\\
 4.75     &       2.7   & &     0.07   & From \cite{1986ApJ...310..561M}\\
 7.5      &       2.6   & &     0.07   & From \cite{1990ApJ...355..102K} \\
 7.5      &       2.64   & &    0.06   & From \cite{1992ApJ...396....3L} \\
 10      &       2.62   & &    0.058   & From \cite{1988ApJ...325....1K} \\
 10.7      &      2.730    & &    0.014   & From \cite{1996ApJ...458..407S} \\
 24.8      &      2.783    & &    0.089   & From \cite{1987ApJ...313L...1J}\\
 33     &        2.81   & &    0.12   & From \cite{1985ApJ...298..710D}\\
 90      &       2.60   & &    0.09   & From \cite{1989ApJ...339..632B} \\
 90       &      2.712   & &    0.020  & From \cite{Schuster1993}\\
\hline
 & (a) & (b) &  & From TRIS \cite{2008ApJ...688...24G} (e) \\
\hline
0.60   &    2.837    & 2.581 &     0.145     & " \\
 0.82   &    2.803    & 2.695 &     0.369     & "  \\
 2.5    &    2.516    & 2.511  &     0.316     & " \\
\hline
\end{tabular}
}
\quad
\scalebox{0.8}{
\begin{tabular}{@{}|c|c|c|c|c|@{}}
\hline
$\nu$ (GHz) & $T$ (K) & $T$ (K) &  $1 \sigma$ error (K)& References \\
\hline
         &        & &     & From compilation in Table 1 of \cite{2008ApJ...688...24G}\\
\hline
     3.7    &     2.59   & &   0.13  &  From \cite{1988ApJ...329..556D} \\
     4.75   &     2.71   & &   0.2  &   From \cite{1984PhRvD..29.2680M}  \\
     2.5 
      &      2.62   & &   0.25  & From \cite{1984PhRvD..29.2686S} \\
     2.5 
      &      2.79   & &   0.15  & From \cite{1986ApJ...311..418S} \\
     2.5 
      &     2.5    & &   0.34  & From \cite{1991ApJ...378..550S} \\
     3.8   &      2.56   & &   0.08  & From \cite{1990ApJ...359..219D} \\
     3.8    &     2.71   & &   0.07  & " \\
\hline
&&&&  From FIRAS \cite{1990ApJ...354L..37M} (figure 3),\\
&&&& with recalibration in \cite{2009ApJ...707..916F}\\
\hline
33   &   2.71548   & &   0.060    &  "  \\
       43   &   2.73548   & &   0.044    & "  \\
       50    &  2.73048    & &  0.033    & "  \\
       58   &   2.72548   & &   0.022    & "  \\
       67    &  2.72548    & &  0.016    & "  \\
       \hline
&&&& From FIRAS \cite{1996ApJ...473..576F}, with\\
       &&&& recalibration in \cite{2009ApJ...707..916F}\\
       \hline
68.1     &   2.72552   & &   0.00011    & \\
  81.5    &    2.72553   & &   0.00011    &  " \\
  95.3    &    2.72555   & &   0.00011    &  " \\
 108.8    &    2.72549    & &  0.00009    &  " \\
 122.3    &    2.72554   & &   0.00007    &  " \\
 136.1    &    2.72540   & &   0.00006    &  " \\
 149.6    &    2.72540   & &   0.00005    &  " \\
 163.4     &   2.72546   & &   0.00004    &  " \\
 176.9    &    2.72555   & &   0.00004    &  " \\
 190.4    &    2.72549   & &   0.00003    &  " \\
 204.2    &    2.72548   & &   0.00003    &  " \\
 217.6    &    2.72551   & &   0.00002    &  " \\
 231.1    &    2.72543   & &   0.00002    &  " \\
 244.9     &   2.72550   & &   0.00002    &  " \\
 258.4    &    2.72550   & &   0.00002    &  " \\
 272.2    &    2.72543   & &   0.00003    &  " \\
 285.7     &   2.72550   & &   0.00003    &  " \\
 299.2    &    2.72551   & &   0.00004    &  " \\
 313.0    &    2.72551   & &   0.00005    &  " \\
 326.5     &   2.72540    & &  0.00006    &  " \\
 340.0    &    2.72534   & &   0.00007    &  " \\
 353.8    &    2.72568   & &   0.00008    &  " \\
 367.2    &    2.72550   & &   0.00008    &  " \\
 381.0     &   2.72546   & &   0.00009    &  " \\
 394.5     &   2.72551   & &   0.00010    &  " \\
 408.0    &    2.72540   & &   0.00010    &  " \\
 421.8    &    2.72551   & &   0.00011    &  " \\
 435.3     &   2.72564   & &   0.00012    &  " \\
 448.8     &   2.72540   & &   0.00013    &  " \\
 462.6    &    2.72533   & &   0.00015    &  " \\
 476.1    &    2.72555   & &   0.00019    &  " \\
 489.9    &    2.72555   & &   0.00023    &  " \\
 503.4    &    2.72564   & &   0.00030    &  " \\
 516.8     &   2.72505   & &   0.00037    &  " \\
 530.6    &    2.72557   & &   0.00045    &  " \\
 544.1     &   2.72593   & &   0.00055    &  " \\
 557.9    &    2.72496   & &   0.00066    &  " \\
 571.4    &    2.72534   & &   0.00080    &  " \\
 584.9    &    2.72569   & &   0.00108    &  " \\
 598.7    &    2.72628   & &   0.00168    &  " \\
 612.2    &    2.72750   & &   0.00311    &  " \\
 625.7    &    2.72064   & &   0.00652    &  " \\
 639.5    &    2.70382   & &   0.01468    &  " \\
\hline
& (a) & (b) &  & From \cite{Dowell:2018}, subtracting the global contribution \\
 &&&& by unresolved extragalactic radio sources \\
\hline
     0.04     &         4317  & 3874 &      963 &  " \\
     0.05       &       2645   & 2405 &     526 &  " \\
     0.06     &         1880   & 1735 &     365 &  " \\
     0.07     &         1189   & 1094 &     208 &  " \\
     0.08     &         969   & 903 &     112 &  " \\
\botrule
\end{tabular}
}
\caption{Adopted data compilation for the cosmic background in terms of equivalent thermodynamic (absolute) temperature. 
Data are collected as described in Section~\ref{data}. 
(a): model by \cite{2008ApJ...682..223G} to subtract the global contribution by unresolved extragalactic radio sources.
(b): model by \cite{2008ApJ...682..223G} to subtract the global contribution by unresolved extragalactic radio sources, but amplified by a factor 1.3 (in terms of antenna temperature) 
to account for possible higher differential number counts at faint flux densities.
(c): for simplicity we report the average of the positive and negative errors (see references for asymmetric errors, where relevant). 
(d): one more digit is shown in corresponding $T$ at columns (b) and (c) to appreciate their little differences.
 (e): we add statistic and systematic errors in quadrature.
}
\label{tab:data}
\end{table*}

\end{document}